\documentclass[12pt]{article}
\usepackage{amssymb}
\usepackage{epsfig}
\def\np#1#2#3{Nucl.\ Phys.\ B#1 (19#3) #2}
\def\npp#1#2#3{Nucl.\ Phys.\ B#1 (20#3) #2}
\def\pl#1#2#3{Phys.\ Lett.\ #1B (19#3) #2}
\def\pll#1#2#3{Phys.\ Lett.\ #1B (20#3) #2}
\def\prb#1#2#3{Phys.\ Rev.\ B #1 (19#3) #2}
\def\prep#1#2#3{Phys.\ Rep.\ #1 (19#3) #2}

\def\frac#1#2{ {{#1} \over {#2} }}

\def\re#1{(\ref{#1})}
\def\beq{\begin{equation}}
\def\eeq{\end{equation}}
\def\beeq{\begin{eqnarray}}
\def\beeqn{\begin{eqnarray*}}
\def\eeeq{\end{eqnarray}}
\def\eeeqn{\end{eqnarray*}}
\def\nome#1{{\label{#1}}}

\def\O{\Omega}
\def\m{\mu}
\def\n{\nu}

\def\s{\sigma}
\def\S{\Sigma}
\def\G{\Gamma}
\def\eps{\epsilon}
\def\L{\Lambda}
\def\l{\lambda}
\def\g{\gamma}

\def\bc{\bar c}

\def\de{\delta}
\def\t,#1{t^{#1}}
\def\f,#1#2#3{f^{#1 #2 #3}} 

\def\sint{S_{\mbox{\footnotesize{int}}}}
\def\sbrs{S_{\mbox{\footnotesize{BRS}}}}

\def\UV{\L_0\to\infty\;}
\def\p{\partial}

\def\d4#1{\frac {d^4 {#1} }{(2\pi)^4}}
\begin{document}
\begin{titlepage}
\begin{flushright}
UPRF-2001-07 \\
\end{flushright}
\vspace{.4in}
\begin{center}
{\large{\bf  Background field method in the Wilson formulation }}
\bigskip \\ M. Bonini and E. Tricarico
\\
\vspace{\baselineskip}
{\small Universit\`a degli Studi di Parma \\ and\\
I.N.F.N., Gruppo collegato di Parma, 
\\viale delle Scienze, 
43100 Parma, Italy} \\
\mbox{} \\
\vspace{.5in}
{\bf Abstract} 
\bigskip 
\end{center} 
\setcounter{page}{0}
A cutoff regularization for a pure Yang-Mills theory is implemented
within the background field method keeping explicit the gauge invariance of
the effective action. The method has been applied to compute the beta 
function at one loop order.
\noindent

\noindent



\end{titlepage}

\section{Introduction} 
The Wilson, or exact, renormalization group (RG) approach regards an 
interacting field theory  as an effective theory, i.e. 
the higher modes of the fields, with respect to some scale $\Lambda$,
generate effective interactions for the lower modes \cite{W}.  
This division of momenta conflicts with gauge invariance  as it
is easy to see in the case of an homogeneous gauge transformation acting 
on some matter field $\phi(x)$:\\
\nonumber 
\beq \phi (x)\to \O (x) \phi (x). 
\eeq 
In the momentum space the gauge transformed field is mapped into a 
convolution with the gauge transformation and then any division of
momenta is lost.
From the functional point of view this conflict appears as a breaking
term in the Slavnov-Taylor (ST) identities for the Wilsonian effective
action \cite{B,noi}, which originates from the introduction of a
cutoff function into the propagators. Although one can rewrite this as
{\it modified} Slavnov-Taylor \cite{Ellw,MT}, it is necessary to prove
that the physical effective action satisfies the ST identities.
As shown in refs. \cite{B,BT} this can be achieved, in perturbation theory, 
by properly
fixing the boundary conditions for the non-invariant couplings in the
Wilsonian effective action at the ultraviolet (UV) scale $\L_0$. 
This is the so-called fine tuning procedure and it is equivalent to
solve the modified Slavnov-Taylor identities at the scale $\L_0$.  

It has been proved \cite{Abbot} for a pure Yang Mills theory that by using a
covariant gauge fixing function depending on some classical external
field, named the {\it background field}, it is possible to define a
gauge invariant effective action.  In \cite{Abbot}
explicit calculations at two loops order have been performed using the
dimensional regularization.  A definite proof that the S-matrix elements 
can be obtained from the background gauge invariant effective action
has been recently established \cite{BC}.

These results have led to apply this background field method to the RG
formulation, with the aim of keeping the gauge invariance explicit
\cite{RW-FLP}. To implement this requirement one allows the cutoff
function to depend on the background field in a covariant way.
However, it is easy to realize, by simple one loop computations, that
the cutoff vertices are not well regularized. A possible way out
consists in adding a mass term for the quantum gauge fields and for
the ghost \cite{becchi00}. 
In this way the BRS invariance is lost and the need of
restoring the ST identities requires a fine tuning procedure.
Therefore the advantages coming from the background gauge method 
are partially lost when the cutoff regularization is used.

The aim of this paper is to show, by an explicit example, how the
cutoff regularization can be successfully implemented within the
background field method.  In section 2 we give the notation. In
section 3 we work out the cutoff regularization preserving the
explicit gauge invariance and we determine the Feynman rules needed to
compute the one-loop two point background amplitude.  In section 4 we
show the transversality of this amplitude and we compute the beta
function at one loop order.
The detailed calculation of the background field wave function 
renormalization is given in the Appendix. Section 5 contains some 
remarks and final comments.

\section{The Background Field}
The gauge-fixed classical action for a pure Yang-Mills theory is given by
\beq\nome{S}
S = S_{cl} + S_{gf} + S_{FP},
\eeq
where 
$S_{cl}= -\frac{1}{4}{\int d^4 x F_{\m \n}^{a} F_{\m \n}^{a}}$ 
is the classical action,
$F_{\m\n}^{a} = \partial _{\m} A_{\n} ^{a}- \partial _{\n} A_{\m}^{a} + g
f^{abc}A_{\m} ^{b} A_{\n} ^{c}$ 
is field strength tensor and $f^{abc}$ are the structure
constants of the group.  
As it has been proposed in \cite{Abbot}, one can choose the gauge fixing 
term  depending on a non-dynamical field \footnote{We set the gauge fixing 
parameter $\alpha=1$ corresponding to the Feynman gauge.}
\beq
\nome{gf} 
S_{gf} = -\frac {1}{2}\int d^4 x \; (\bar D _{\m} ^{ab} (A_{\m} ^b - 
{\bar A}_{\m} ^b ))^2,
\eeq
where $\bar A_{\m}^a$ is the {\it{background field}} and 
the covariant derivative is
\beq
\bar D _{\m} ^{ab} = \partial _{\m} \de ^{ab} + g f^{acb} \bar A_{\m} ^c.
\eeq
As a consequence of \re{gf} the ghost term also depends on the background field
and reads
\beq
S_{FP} = - \int d^4 x \; \bar c^a \bar D _{\m} ^{ab} D _{\m} ^{bd} c^d,
\eeq
with $D _{\m} ^{ab} = \partial _{\m} \de ^{ab} + g f^{acb}  A_{\m} ^c$.
The particular choice of the gauge fixing term makes the action \re{S} 
invariant under the following {\it background gauge transformation}:
\beq\nome{gaugetr}
\de A_{\m} ^a = D_{\m} ^{ab} \l ^a,\;\;\;\;\;
\de \bar A_{\m} ^a =\bar  D_{\m} ^{ab} \l ^a,
\eeq
\beq\nome{brs}
\de c^a = g f^{abc} c^b \l ^c,\;\;\;\;\;
\de \bar c^a = g f^{abc} \bar c ^b \l ^c,
\eeq
where $\l=\l(x)$ is the infinitesimal gauge parameter.
The invariance of the classical action under the gauge transformation  of 
the field $A_\m$ is implemented at the quantum level by the BRS symmetry:
\beq\nome{debrs}
\de A_{\m} ^a =\eta D_{\m} ^{ab} c^b,\;\;\;\;\;
\de c^a = -\frac12\eta g f^{abc} c^b c^c,\;\;\;\;\;
\de \bar c^a =\eta \bar D_\m (A-\bar A)_\m^a
\eeq
where $\eta$ is a Grassmann parameter.
Adding to the action \re{S} the source term associated to the BRS 
variations \re{debrs} of $A_\m$ and $c$ one has
\beq\nome{SYM}
\sbrs[\Phi,\g]=S+\int d^4x ( u_\m^a D_{\m}^{ab} c^b
-\frac g2 f^{abc}v^a c^b c^c)\,,
\eeq
where we have denoted by $\Phi= (A_{\m} ^a, c^a, \bar c^a)$ 
and $\g=(u_\m^a, v^a)$  the fields and the BRS sources.

In the conventional functional approach one defines the generating functional
\beq
\nome{funct}
Z[J,\g] = e^{iW} = \int \it D \Phi\;\; e^{iS_{BRS}[\Phi, \g] + i(J \Phi)},
\eeq
where $J=(j_{\m}^a, \bar\eta^a,  -\eta^a)$ are the field sources and we 
introduced the notation 
\beq
(J,\Phi)=\int d^4x\bigl(j_\m^a A_\m^a+\bar\eta^a c^a+ \bar c^a \eta^a\bigr).
\eeq
The symmetry of the BRS action with respect to the background gauge 
transformation \re{gaugetr} translates for the effective action 
\beq
\G[\Phi _{cl},\g] = W[J,\g] - (J\Phi _{cl}),\;\;\;\;\;
\Phi_{cl} = \frac {\de W}{\de J},
\eeq
into
\beq\nome{ward}
{\it G^a} \G[\Phi_{cl},\g] + {\it \bar G ^a} \G[\Phi_{cl},\g] = 0,
\eeq
where 
\beq
{\it G^a} = D_{\m} ^{ab} \frac {\de}{\de A_{{\m},cl} ^b} + 
g f^{abd} c_{cl} ^b \frac{\de}{\de c_{cl} ^d} +g f^{abd} \bar c_{cl} ^b 
\frac{\de}{\de \bar c_{cl} ^d}
\eeq
and
\beq
{\it \bar G ^a} = \bar{D}_{\m} ^{ab} 
\frac {\de}{\de \bar A ^b _{{\m},cl} }.
\eeq
Similarly the BRS transformation translates into the ST identities.
The Ward identity \re{ward} expresses the gauge invariance of the
effective action $\G[\Phi _{cl},\g]$ for $\bar A_{\m} ^a =A_{\m cl}^a$.  

In the generating functional \re{funct} one can set 
$Q_{\m} ^a = A_{\m} ^a - \bar A _{\m} ^a$ 
obtaining the  new functional $\tilde W$ related to $W$ by
\beq
\tilde W[J,\g;\bar A] = W[J,\g] -\int d^4x j_\m ^a \bar A_\m ^a
\eeq
and the corresponding effective actions are related by
\beq
\nome{gi}
\tilde \G [\tilde \Phi _{cl},\g; \bar A] = \G[\Phi _{cl},\g],\;\;\;\;
\tilde{A}_{\m cl}^a = A_{\m cl} ^a - \bar A_{\m} ^a.
\eeq
From this relation it is evident that the gauge invariant effective action
$\G[\Phi _{cl},\g]_{\bar A =A_{cl}}$  can be obtained from 
$\tilde \G [\tilde \Phi _{cl},\g; \bar A]_{\tilde{A}_{cl} =0}$, i.e.
by evaluating 1PI Green functions with background fields on the 
external legs and the quantum fields $Q$, $c$ and $\bar c$ inside loops.
A rigorous proof of this {\it {background gauge equivalence}}
has been recently given in \cite{BC}.

One has to realize that the derivation of the Ward identities
\re{ward} is purely formal since the regularization procedure has not
been taken into account. If one uses the dimensional regularization
all the symmetries are preserved (for a pure YM theory) and the Ward
identity also holds for the renormalized background effective action.
In particular one can derive the $\beta$-function from the two point 
background amplitude, since the gauge coupling and wave function 
renormalizations are related \cite{Abbot}. 

In the following sections we will use the cutoff regularization, by adjusting 
the Wilson-Polchinski RG approach to the background field method.
To maintain the background
gauge invariance we will introduce a covariant cutoff function as proposed
in \cite{RW-FLP}. Moreover we will be forced to add a mass term 
to the bare action, to properly regularize the one loop contribution.

\section {Covariant regularization}
In the following we will specify to the SU(2) case and, to simplify 
the notation, we will use the dot and wedge SU(2) products.
The BRS action $\sbrs$ expressed in terms of the quantum gauge field $Q$ is 
\beq
\nome{s}
S=\int _{x} \Bigl\{-\frac14 \bar F_{\m\n} \cdot  \bar F_{\m\n} +
\frac{1}{2}Q_\m\cdot\bar{D}^{2} Q_\m
+ Q_\m\cdot \bar F_{\m \n}\wedge Q_{\n} - \bar c\cdot \bar{D}^{2} c\Bigr\} 
+ \sint[Q,c,\bar A;\g],
\eeq
where $\bar F_{\m\n}^a$ is the field strength tensor of the background field
and
we have explicitly written the terms quadratic in the quantum
field $Q_\m$ and in the ghosts. The remaining terms, which have been collected
in $\sint$, do not contribute to the one-loop 
vertex functions with background fields on the external legs and for this
reason it is not necessary to work out their expression.
Notice that the field $Q$ transforms according to the adjoint representation
under the background gauge transformation \re{gaugetr}.

To select the modes of the quantum fields below the UV cutoff
$\Lambda_0$ we introduce a cutoff function $K_{\Lambda_0}$ and make
the following change of variables in the generating functional $\tilde W$
\beq\nome{redef}
Q_\m \to K_{\L_0} Q_\m\,, \;\;\;\;\;\;\;\; c\to K_{\L_0} c\,,
\eeq
(it is not necessary to introduce a new $\bc$ field). 
The invariance of the action with respect to the background gauge 
transformation \re{gaugetr} can be maintained if the regularized fields 
transform according to the adjoint representation.
This can be achieved if $K$ is a function of an
appropriate covariant operator, such us $\bar D^2$ and then we choose
\beq \nome{k} 
K_{\L_0}\equiv K\left[-{\bar D}^2 / \L_0^2
\right]=\sum_{n=0}^{n=\infty} \frac{1}{n!}
K^{(n)}(0){\left({\frac{-\bar D}{\L_0}}\right)^n}=K(-\p^2/\L_0^2)+ \cdots\,,
\eeq 
where $K^{(n)}(0)=\left[\frac{d^nK(x)}{dx^n}\right]_{x=0}$ 
and the dots refer to the terms containing the $\bar A$ field.
After the substitution \re{redef},  the gauge propagator is multiplied 
by the factor $K(p^2/\L^2_0)^{-2}$ while the ghost one by the factor
$K(p^2/\L^2_0)^{-1}$.
As usual in the Wilson RG formulation, 
one chooses the cutoff function such as to suppress  the propagation 
of the modes with $p^2>\L^2_0$.
However, the choice \re{k} of the cutoff function produces new interactions 
among the quantum and the background fields which are multiplied 
by the cutoff function $K(p^2/\L_0^2)$ (or its derivatives, see later).
Thus the loop momenta which are
suppressed by the inverse of the cutoff function in the propagator are enhanced
by the cutoff function in these new  vertices and the regularization fails.
This fact is not surprising and is essentially 
a consequence of the Ward identity, which relates the vertex with the
inverse of the propagator\footnote
{We thanks Prof. Carlo Becchi for fruitful discussions on this
point.}.
To overcome this obstacle we introduce in the action a
mass terms for the quantum fields
\beq\nome{mass}
S_m=\int _{x} \biggl[-\frac{1}{2}Q_{\m} \cdot M_{Q} Q_{\m}+ 
\bar c \cdot M_{c}\ c\biggr]\,,
\eeq
where the matrices
$M_{Q}$ and $M_{c}$ depend on the cutoff function and must satisfy 
the requirement that this mass term does not 
generate relevant interactions in the $\UV$ limit. For instance, 
by choosing the exponential covariant cutoff function:
\beq\nome{expcut}
K\left[-{\bar D}^2 /  \L_0^2 \right] = 
e^{\frac{{\bar D}^2}{2 \L_0^2}}\,,
\eeq
the matrices
$M_{Q}$ and $M_{c}$ have the structure
\beq
M_{Q}^{ab}=\L_0^2\left(\de^{ab}-(K^2)^{ab}+K^{ac}
{\frac{\left(\bar{D}^{2}\right)^{cd}}{\L_0^2}}
K^{db}\right),\;\;
M_{c}^{ab}=2\L_0^2\left(\de^{ab}-K^{ab}+ 
{\frac{\left(\bar{D}^{2}\right)^{ac}}{2 \L_0^2}}
K^{cb}\right).
\eeq
Notice that this structure holds for every cutoff function satisfying the 
conditions 
\beq\nome{cond}
K(0)=1\,\;\;\;\;\;\;\;
K'(0)=-1/2\,. 
\eeq

After using \re{s}, \re{redef} and \re{mass}, the regularized action becomes
\beeq\nome{actreg}
&&S_{\L_0}=\int_x \Bigl[-\frac14 \bar F_{\m\n}\cdot \bar F_{\m\n} - 
\frac12 \L_0 ^2 Q_{\m} \cdot(1-K^2) Q_{\m} -
             g \bar F_{\m \n}\cdot(K Q_{\m})\wedge(K Q_{\n}) 
\nonumber\\
&&\phantom{\int_x }
           +2 \L_0 ^2 \bar c \cdot \left(1 - K \right)  c +
            \sint[KQ,Kc,\bar A;\g]\Bigr]\,.
\eeeq
This action preserves the background gauge invariance \re{gaugetr} but breaks 
the BRS symmetry \re{brs} and a fine tuning procedure must be imposed in order 
to restore the ST identities. 
This analysis can be performed as in the standard Wilsonian approach
for gauge theories \cite{B,noi}. 
First one introduces a cutoff dependent BRS transformation,
studies the modified ST identities and determines the non-invariant
couplings which compensate the breaking introduced by the cutoff.
From the quantum action principle these couplings are order $\hbar$  
and therefore affect the background field amplitudes starting from the second 
loop order. 
In this paper we are only interested in one-loop computations and therefore
this fine-tuning problem can be ignored. However, this procedure 
is unavoidable in the complete analysis.

From \re{actreg} the gauge and ghost propagators read
\beq
\mathcal{D}^{ab}_{\m \n}(q)={\frac{\de^{ab}g_{\m\n}}{\L_0^2\left(1-K^2(q)
\right)
}}\,,\;\;\;
\mathcal{D}^{ab}(q) = - {\frac{\de^{ab}}{2 \L_0^2\left(1-K(q)\right)}}\,,
\eeq
where
\beq
K(q)\equiv K(q^2/\L_0^2)=e^{-\frac{q^2}{2\L_0^2}}\,.
\eeq
Notice that for large $q$  (i.e. $q>>\L_0$) these propagators become constant, 
and the UV finiteness of the loop integral will be ensured by the cutoff 
function in the vertices.  

In order to evaluate the Feynman rules coming from \re{actreg}, 
we first expand  $\left(KQ\right)^{a}_{\m}$ in terms of the $\bar A$ 
field. For instance the terms with one and two $\bar A$ fields are given by
\beq\nome{ka1}
\left(KQ\right)^{a}_{\m}|_{1 \bar A} =
\sum _{n=1}^{\infty} \frac{1}{n!2^n\L_0^{2n}}
\sum _{k=0}^{n-1}\left(\partial^2\right)^k \Delta_{\bf {1}}^{ab}
\left(\partial^2\right)^{n-k-1}Q_{\m}^{b}
\eeq
and
\beeq\nome{ka2}&&
\left(KQ\right)^{a}_{\m}|_{2 \bar A} =
\sum_{n=1}^{\infty} \frac{1}{n!2^n\L_0^{2n}}
\sum _{k=0}^{n-1}\left(\p^2\right)^k \Delta_{\bf {2}}^{ab} 
\left(\p^2\right)^{n-k-1} Q_{\m}^{b}\nonumber\\
&&\phantom{\left(KQ\right)^{a}_{\m}|_{2 \bar A}}
+\sum_{n=2}^{\infty}\frac{1}{n!2^n\L_0 ^{2n}}
\sum_{k=0}^{n-2}\;\sum_{l=0}^{n-k-2}
\left(\p^2\right)^k \Delta_{\bf {1}}^{ac} 
\left(\p^2\right)^l \Delta_{\bf {1}}^{cb}
\left(\p^2\right)^{n-k-l-2} Q_{\m}^{b},
\eeeq
where
\beq
\Delta _{\bf 1} ^{ab} = g \eps^{acb}\left(\bar A_{\rho}^{c}
\partial _{\rho} + \partial_{\rho} \bar A_{\rho}^c\right)
\eeq
and
\beq
\Delta _{\bf 2} ^{ab} = g^2 \eps^{ace} \eps^{edb} 
\bar A_{\rho}^{c} \bar A_{\rho}^d.
\eeq

In the following section we will compute the one loop two point function for 
the background field. For this purposes we need to evaluate the Feynman
rules with at most two $\bar A_\m$ fields.

By inserting \re{ka1} in \re{actreg} the 
$Q_\m^a(q)$-$\bar A_{\m_1}^{a_1}(p_1)$-$Q_\n^b(p)$-vertex is given by
\beeq\nome{v3a}&&
V_{\m\m_1\n}^{aa_1b}(q,p_1,p)=ig\eps^{aa_1b}\left[
2K(q)K(p)\,(g_{\m_1\n}p_{1\m}-g_{\m_1\m}p_{1\n})
\right.\nonumber
\\
&&\phantom{V_{\m\m_1\n}^{aa_1b}(q,p_1,p)=ig\eps^{aa_1b}}
\left.
+\L_0^2\,(K(q)+K(p))F(q,p)
g_{\m\n}(q-p)_{\m_1}
\right]
\eeeq
where 
$$
F(q,p)=(K(q)-K(p))/(q^2-p^2)\,.
$$

The vertex with two $Q_\m$ and two $\bar A_\m$ fields receives
contribution from the term $\bar F_{\m\n}\cdot KQ_\m\wedge KQ_\n$
in \re{actreg} and from the covariant cutoff functions \re{ka2}.
We do not need to compute the former since it does not contribute to the 
one loop two point functions (i.e. to the tadpole diagram). 
The remaining terms of the
$Q_\m^a(q)$-$\bar A_{\m_1}^{a_1}(p_1)$-$\bar A_{\m_2}^{a_2}(p_2)$-$Q_\n^b(p)$-vertex are
\beeq\nome{v4a}
&&
V_{\m\m_1\m_2\n}^{aa_1a_2b}(q,p_1,p_2,p)=g^2\L^2g_{\m\n}(\de^{a_1a_2}\de^{ab}-
\de^{a_1b}\de^{a_2a})
\biggl\{(K(q)+K(p))F(q,p)
g_{\m_1\m_2}
\nonumber\\&&
\phantom{V_{\m\m_1\m_2\n}^{aa_1a_2b}}
-\biggl[\frac{F(q,p)-F(q,p+p_2)}{p^2-(p+p_2)^2}\,K(q)+
\frac{F(q,p)-F(p,q+p_1)}{q^2-(q+p_1)^2}\,K(p)
\nonumber\\&&
\phantom{V_{\m\m_1\m_2\n}^{aa_1a_2b}}
+F(q,q+p_1)F(p,p+p_2)\biggr]
(2q+p_1)_{\m_1}(2p+p_2)_{\m_2}
\biggr\}
\;\; + 1\leftrightarrow 2\,.
\eeeq

The interactions of the ghosts with the background fields  can be 
obtained from \re{actreg} and expanding $K(\bar D^2/\L_0^2) c$ in powers of
$\bar A_\m$ as we have done for $K\, Q_\m$.
The $\bc^a(q)$-$\bar A_{\m_1}^{a_1}(p_1)$-$c^b(p)$-vertex is given by
\beq\nome{vgh1a}
V_{\m_1}^{aa_1b}(q,p_1,p)=
-2i\L_0^2g \eps^{aa_1b}(q-p)_{\m_1}F(q,p).
\eeq

The 
$\bc^a(q)$-$\bar A_{\m_1}^{a_1}(p_1)$-$\bar A_{\m_2}^{a_2}(p_2)$-$c^b(p)$-vertex
is given by
\beeq\nome{vgh2a}
V_{\m_1\m_2}^{aa_1a_2b}(q,p_1,p_2,p) &=&
2\L_0^2g^2\biggl\{
\left(\de^{ab}\de^{a_1a_2}-\de^{a_1b} \de^{a_2a}\right)
\biggl[(2q+p_1)_{\m_1}(2p+p_2)_{\m_2}
\nonumber\\&&
\times\frac{F(q,p)-F(q,p+p_2)}{p^2-(p+p_2)^2}
-g_{\m_1\m_2}F(q,p)\biggr]+1\leftrightarrow 2 \biggr\}.
\eeeq
Though the vertices \re{v3a}-\re{vgh2a} have been computed
using the cutoff function \re{expcut}, their expression in term of the
functions $K(q)$ and $F(q,p)$ holds for every cutoff function
satisfying \re{cond}.

From the above expression for the vertices and the propagators, it is
clear that the UV finiteness of the loop integrals is ensured if the
function $K(q^2/\L^2)$ decreases rapidly enough in the region
$q^2>>\L^2$.

\section{One loop computations} 

As briefly discussed in Section 2, one is only interested to discuss
vertices with background external fields. We will explicitly compute
the two point amplitude for the background field and we will verify
that is transverse, as it must be since the regularization preserves
the background gauge invariance.
There are four Feynman graphs contributing to this amplitude, 
which are depicted in figure 1a-1d.
\begin{figure}[htbp]
\begin{center}
\epsfysize=6cm
\epsfbox{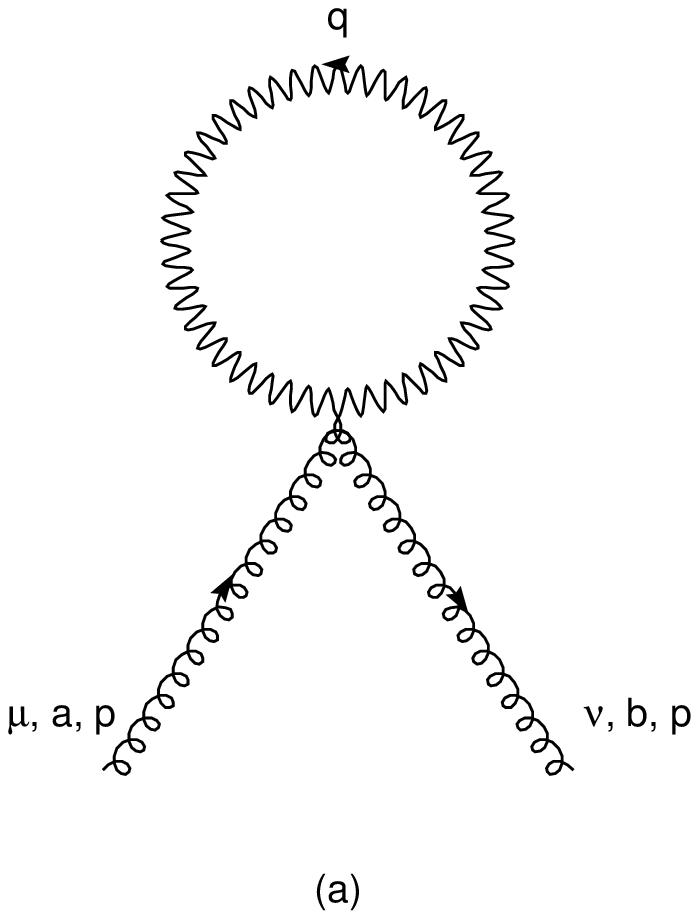}
\epsfysize=4.5cm
\epsfbox{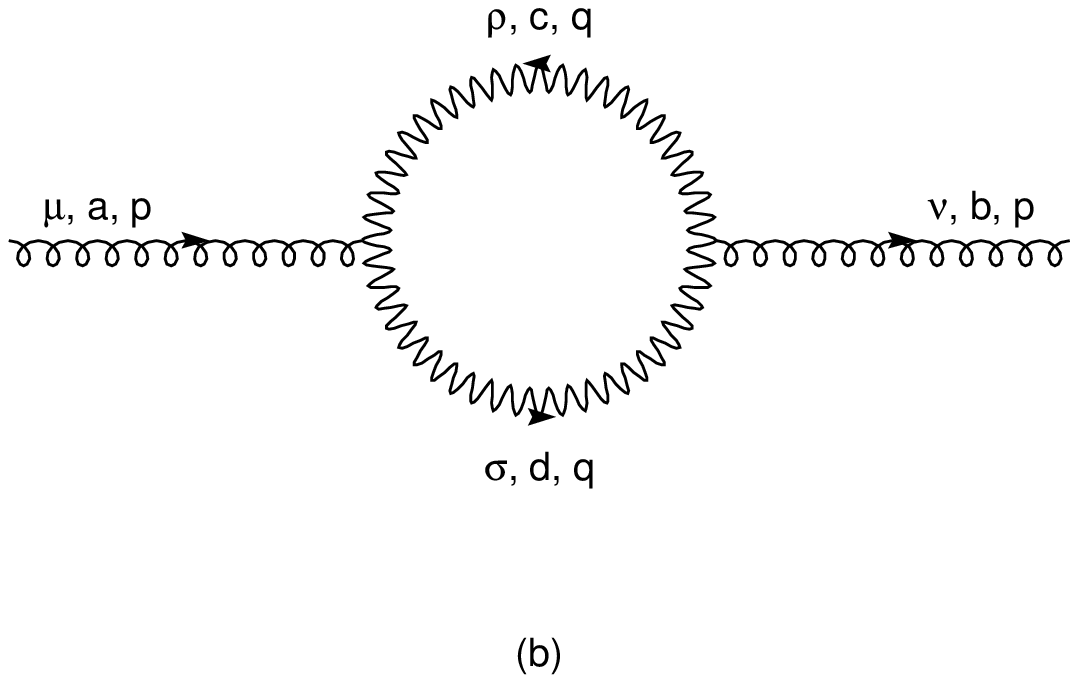}
\end{center}
\begin{center}
\epsfysize=6cm
\epsfbox{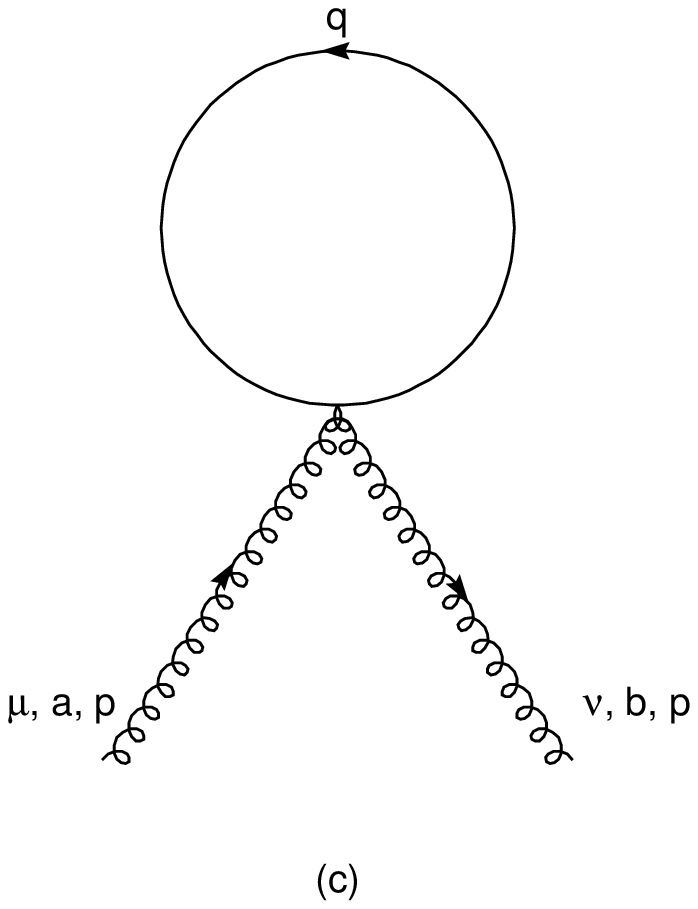}
\epsfysize=4.5cm
\epsfbox{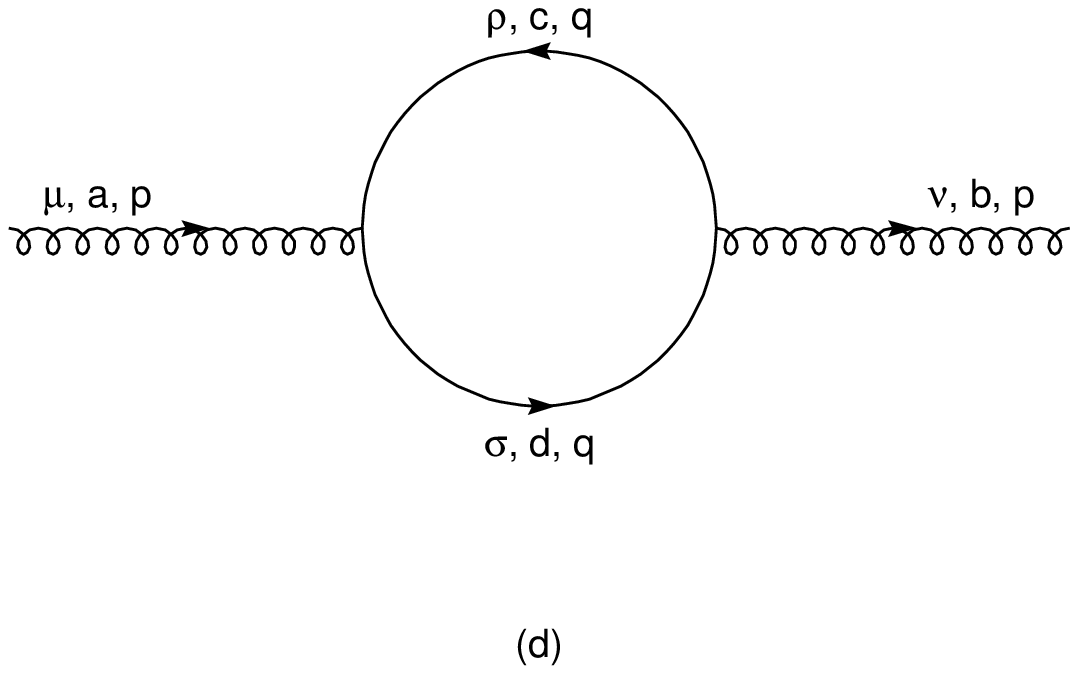}
\end{center}
\caption{\small{Graphical contribution to the two point function
with background external legs which are depicted as curly lines.
The wavy and the full lines in the loops refer to the 
quantum gauge and ghost field, respectively.
}}
\end{figure}
The corresponding loop integrals are
\beeq\nome{4a}
&&
{\cal G}^{[1a]ab}_{\m\n}(p;\L_0)= g^2\de^{ab}\int_q\Biggl\{16K(q)F(q,q) 
g_{\m\n}
+\Biggl[(2q+p)_\m\,(2q+p)_\n\biggl(4F^2(q,q+p)
\nonumber\\&&
\phantom{{\cal G}^{[1a]ab}_{\m\n}(p;\L_0)}
+ 8\frac{F(q,q)-F(q,q+p)}{q^2-(q+p)^2}\,K(q)\biggr)+\;p\to-p\Biggr]\Biggr\}
\frac{1}{1-K^2(q)}\,,
\eeeq

\beeq\nome{4b}&&
{\cal G}^{[1b]ab}_{\m\n}(p;\L_0)= g^2\de^{ab}\int_q\biggl\{
4(2q+p)_\m\,(2q+p)_\n F^2(q,q+p)[K(q)+K(q+p)]^2
\nonumber\\&&
\phantom{{\cal G}^{[1a]ab}_{\m\n}(p;\L_0)}
+ 8(g_{\m\n}p^2-p_{\m}p_{\n})\frac{K^2(q)K^2(q+p)}{\L_0^4}\biggr\}
\frac{1}{(1-K^2(q))
(1-K^2(q+p))}\,,
\eeeq

\beeq\nome{4c}&&
{\cal G}^{[1c]ab}_{\m\n}(p;\L_0)=-2g^2\de^{ab}\int_q\Biggl\{2F(q,q) 
g_{\m\n}
+\Biggl[(2q+p)_\m\,(2q+p)_\n\frac{F(q,q)-F(q,q+p)}{q^2-(q+p)^2}
\nonumber\\
&&
\phantom{{\cal G}^{[1c]ab}_{\m\n}(p;\L_0)=2g^2\de^{ab}\int_p\Biggl\{2F(q,q) 
g_{\m\n}}
+\;p\to-p\Biggr]\Biggr\}
\frac{1}{1-K(q)}\,,
\eeeq

\beeq\nome{4d}&&
{\cal G}^{[1d]ab}_{\m\n}(p;\L_0)=-2g^2\de^{ab}\int_q\,
(2q+p)_\m\,(2q+p)_\n \frac{F^2(q,q+p)}{(1-K(q))(1-K(q+p))}\,,
\eeeq
where in the first two
contribution the symmetry factor $1/2$ has been included and  
$$
F(q,q)=\lim_{p\to0}F(q,q+p)=\frac{d}{dq^2}K(q^2/\L_0^2)
\equiv \frac{K'(q)}{\L_0^2}\,.
$$ 
One can easily verify the transversality of the two-point function. 
Indeed form the gauge field loop (fig.1a and 1b) one finds 
\beeq&&
p_\m p_\n( {\cal G}_{\m\n}^{[1a]ab}+ {\cal G}_{\m\n}^{[1b]ab})=
4g^2\de^{ab}\int_q\biggl[-2\frac{K^2(q)-K^2(q+p)}{1-K^2(q)}
+\frac{(K^2(q)-K^2(q+p))^2}{(1-K^2(q))(1-K^2(q+p))}\biggr]\nonumber
\\&&
\phantom{p_\m p_\n( {\cal G}_{\m\n}^{[1a]ab}+ {\cal G}_{\m\n}^{[1b]ab})
}
=8g^2\de^{ab}\int_q\,\frac{K^2(q)-K^2(q+p)}{1-K^2(q)}\,
\biggl[\frac{K^2(q)}{1-K^2(q+p)}-1\biggr]=0\,,
\nonumber
\eeeq
where the last two equalities has been obtained by performing  
the change of the integration variable $q\to -p-q$.
Similarly  from the ghost loop (fig.1c and fig 1d) one finds
$$
p_\m p_\n( {\cal G}_{\m\n}^{[1c]ab}+ {\cal G}_{\m\n}^{[1d]ab})=
2g^2\de^{ab}\int_q\biggl[-2\frac{K(q)-K(q+p)}{1-K(q)}
+\frac{(K(q)-K(q+p))^2}{(1-K(q))(1-K(q+p))}\biggr]=0\,.
$$
Therefore the four graphs \re{4a}-\re{4d} sum up to
$$
{\cal G}_{\m\n}^{ab}(p,\L_0)=\de^{ab}(p^2g_{\m\n}-p_\m p_\n) {\cal G}(p,\L_0)
$$
where
\beq\nome{G}
{\cal G}(p,\L_0)=
\frac1{6p^2}{\cal G}_{\m\m}^{aa}(p,\L_0).
\eeq

The renormalized background two point amplitude 
$$
\G_{\m\n}^{ab}(p)=\de^{ab}(p^2g_{\m\n}-p_\m p_\n) \S(p)\,,
\;\;\;\;\; \;\;\;\;\; \;\;\;\;\; 
\S(p)|_{p^2=\m^2}=0
$$
at one loop level is given by
$$
\S(p)=\lim_{\L_0\to\infty}({\cal G}(p,\L_0)-\s_1(\m/\L_0))\,,
$$
where the relevant coupling $\s_1$ is 
\beq\nome{s1}
\s_1(\m/\L_0))={\cal G}(p,\L_0)|_{p^2=\m^2}
\eeq
and $\m$ is the renormalization scale. This relevant coupling, which
depends on the regularization, will be computed in the appendix
in the case of a polynomial cutoff function.

\subsection{One-loop beta function}
In the Wilsonian approach one can show \cite{BMS} that the beta
function can be obtained from the bare coupling constant $g_B$ by
\beq\nome{betaw}
\beta(g)\equiv\m\p_\m g=\frac{\L_0\p_{\L_0}g_B}{\p_g g_B}\,.
\eeq
Due to the background gauge invariance the bare coupling constant 
$g_B$ is related to the relevant coupling $\s_1$ by
$$
g_B=g(1+\s_1)^{-1/2}
$$ 
from which, at one-loop order, one obtains
\beq\nome{betaw1}
\beta(g)=-\frac12 \L_0\frac{\p}{\p\L_0}\s_1(\m/\L_0)\,.
\eeq
In the Appendix we will compute this coupling by using a polynomial 
cutoff function. 
However the first coefficient of the beta function is
independent of the regularization and therefore it should be computed 
without specifying the cutoff function.

In order to determine the beta function we first consider the 
contribution to \re{G} 
originating from the two diagrams with the gauge field loop 
(see  Figs. 1a and 1b). 
From \re{4a} and \re{4b} one obtains 
\beeq&&\nome{trvec}
{\cal G}_{\m\m}^{[1a]aa}+{\cal G}_{\m\m}^{[1b]aa}
= 16g^2\int_q \Biggl\{
\frac{3p^2K^2(q)K^2(q+p)}{\L_0^4(1-K^2(q))(1-K^2(q+p))} +
\frac{8 K(q) K'(q)}
{\L_0^2 (1-K^2(q))}
\nonumber\\[6pt]&& 
\phantom{{\cal G}(p,\L_0)= \frac{8g^2}{3}\int_q}
-\frac{(2q+p)^2}{\L_0^2\;p\cdot(2q+p)}
\biggl(\frac{K(q) K'(q)}{1-K^2(q)}-\frac{K(q+p) K'(q+p)}{1-K^2(q+p)}
\biggr)\Biggr\}.
\eeeq
Similarly, from the two diagrams with the ghost loop 
(see  Figs. 1c and 1d) one gets
\beeq&&\nome{trgh}
\!\!\!\!\!\!\!\!\!\!{\cal G}_{\m\m}^{[1c]aa}+{\cal G}_{\m\m}^{[1d]aa}
=
-\frac{4g^2}{\L_0^2}\int_q \Biggl\{
\frac{8 K'(q)}{1-K^(q)}
-\frac{(2q+p)^2}{p\cdot(2q+p)}
\biggl(\frac{K'(q)}{1-K(q)}-\frac{K'(q+p)}{1-K(q+p)}
\biggr)\Biggr\}.
\eeeq
Because of the infrared divergence \footnote{Notice that the mass term
\re{mass} does not change the infrared behavior of the propagators.}  
one can not set $p^2=\m^2=0$ in these integrals,
however as far as the beta function is concerned, in \re{betaw} one can take 
$\m/\L_0\to 0$ \cite{BMS}.
Therefore, from \re{G} and \re{betaw1}, the ghost contribution to the 
beta function is 
\beq
-\frac12\,\L_0\frac{\p}{\p\L_0}\biggr(\frac{1}{6\m^2}\biggr(
{\cal G}_{\m\m}^{[1c]aa}+{\cal G}_{\m\m}^{[1d]aa}\biggr)\biggr)\vert_{\m^2=0}
\eeq
and, by Taylor expanding in $p$ the integrand of \re{trgh}, it becomes
\beq\nome{genres}  
\frac13\,\frac{g^2}{16\pi^2}
\int_0^\infty dq^2\,q^2 \L_0\frac{\p}{\p\L_0}
\Biggl\{\frac1{\L_0^4}
\Biggl[k'(x)+3x k''(x)+
\frac23 x^2 k'''(x)\Biggr]
\Biggr\},
\eeq
where $k(x)\equiv [K'(x)/(1-K(x))]$ and $x=q^2/\L_0^2$.
After integrating by parts and using \re{cond},
this contribution is 
$$
-\frac{\,g^2}{16\,\pi^2}\frac23\,.
$$
Performing  a similar analysis for the gauge loop contribution, one obtains
\beeq&&
-\frac12\,\L_0\frac{\p}{\p\L_0}\biggr(\frac{1}{6\m^2}\biggr(
{\cal G}_{\m\m}^{[1a]aa}+{\cal G}_{\m\m}^{[1b]aa}\biggr)\biggr)\vert_{\m^2=0}
\nonumber\\
&&\phantom{-\frac12}
=-\frac43\,\frac{g^2}{16\pi^2}
\int_0^\infty dq^2\,q^2 \L_0\frac{\p}{\p\L_0}
\Biggl\{\frac1{\L_0^4}
\Biggl[\frac{3K^4(x)}{(1-K^2(x))^2}+h'(x)+3x h''(x)+
\frac23 x^2 h'''(x)\Biggr]
\Biggr\}
\nonumber\\
&&\phantom{-\frac12}
=-\frac{20}{3}\,\frac{g^2}{16\pi^2}\,,
\eeeq
with $h(x)\equiv [K(x)K'(x)/(1-K^2(x))]$.
Summing up the two contributions one obtains 
the well-known one-loop result
$$
\beta(g)=-\frac{22}{3}\,\frac{1}{16\pi^2} g^3\,.
$$

\section{Conclusions} 
In this paper we have implemented 
a cutoff regularization which maintains the gauge invariance
explicit. It is based on the introduction of a background field and a
cutoff regulator covariantly depending on this field.  
The finiteness of the loop
integrals is ensured by the presence of a cutoff function which
multiplies all interactions and suppresses the loop momenta in the UV
region. Moreover, by adding a mass term for the quantum fields the
propagators remain finite in this region. This mass term is necessary 
if one wants to keep gauge invariance and, at the same time, to have 
an efficient regularization.  
The only request we made is that this mass term does not 
introduce relevant interactions into the action when removing the cutoff.
The explicit gauge invariance of the effective action 
has been exploited to compute the beta function at one loop order 
from the  wave function renormalization of the background field.
The fact that
this result is independendent of the cutoff function choice 
is a check of the consistency of our computation.

All the background amplitudes can be made finite to all loops by using an
appropriate cutoff function, such as the exponential one \re{expcut}.
The main problem one has to address before extending this analysis to
higher loops lies on the explicit
BRS symmetry breaking introduced by the mass term.  Although this
symmetry only affects the quantum fields, the ST identities are
necessary to show the background gauge equivalence \cite{BC}.
Therefore a complete analysis requires the determination of the
symmetry breaking counterterms. As in the standard (i.e. without the
background field) Wilsonian formulation of gauge theories, these
counterterms can be calculated by introducing a generalized BRS
symmetry, dependent on the cutoff function, and studying the
corresponding ST identities.  Also in this case one can show that by
imposing these modified ST identities, the renormalized effective
action satisfies the ST, in the limit in which the cutoff is removed.
The solution of this fine tuning problem is outside the aim of the
present work and is left to a further publication. We only remarque
that the background gauge invariance greatly simplifies the
determination of the symmetry breaking counterterms. Moreover, in
order to determine background gauge amplitudes only few of these
counterterms are needed.  For instance for computing the second
coefficient of the beta function one only needs to determine the
couplings of the interactions with at most two quantum gauge fields.

\vspace{1cm}\noindent{\large\bf Acknowledgements}
We have greatly benefited  from discussions with Carlo Becchi and 
Giuseppe Marchesini.

\section*{Appendix}
In this appendix we compute the relevant coupling $\s_1$ given in \re{s1}. 
To performe this calculation one needs to specify the cutoff function.
We use the polynomial cutoff function 
\beq\nome{polcut}
K(q)=\frac{1}{\left(1+\frac{q^2}{4\L_0^2}\right)^2},
\eeq 
which satisfies the conditions \re{cond} and allows to evaluate the 
Feynman integrals by exploiting the Feynman parameterization.

We first concentrate on the ghost loop contribution \re{trgh}, which 
becomes
\beeq\nome{gh}
&&\!\!\!\!\!\!\!{\cal G}_{\m\m}^{[1c]aa}+{\cal G}_{\m\m}^{[1d]aa}
\nonumber\\
&&\!\!\!\!\!\!\!
=128\;g^2\L_0^4\int_q
\biggl\{\frac{8}{q^2(q^2+4\L_0^2)
(q^2+8\L_0^2)}
-(2q+p)^2\biggl[\frac{1}{q^2(q+p)^2\;
(q^2+4\L_0^2)(q^2+8\L_0^2)}
\nonumber\\
&&\!\!\!\!\!\!\!
+\frac{1}{q^2(q^2+4\L_0^2)(q^2+8\L_0^2)
((q+p)^2+8\L_0^2)}
+\frac{1}{(q^2+4\L_0^2)(q^2+8\L_0^2)
(q+p)^2((q+p)^2+4\L_0^2)}
\biggr]\biggr\}\nonumber.
\eeeq
One can easily see that the $q$-integral is UV. 
The first term in the integrand is independent of $p$ and can be easily 
evaluated obtaining
\beq
1024g^2\L_0^4\int_q \frac{1}{q^2(4\L_0^2+q^2)(8\L_0^2+q^2)}=
\frac{16\L_0^2}{\pi^2}g^2 \log\,2.
\eeq
For the remaining terms one uses the Feynman parameterization. 
For instance one has 
\beeq
\int_q\frac{(2q+p)^2}{q^2(4\;\L_0^2+q^2)(8\;\L_0^2+q^2)(q+p)^2}=
3!\int_q \int_{x,y,z}\;
\frac{[2q+p(1-2z)]^2}{\biggl[q^2+z\;(1-z)\;p^2+4\;\L_0^2(x+2y)\biggr]^4},
\eeeq
where $\int_{x,y,z}\equiv \int_0^1\;dz \int_0^{1-z}\;dy\int_0^{1-x-y}\;dx$.
After integrating with respect to $q$ and the Feynman parameters the ghost 
loop contribution gives
\beq
-\frac{p^2}{4\pi^2}\log\,\frac{p^2}{\L_0^2}+
\frac{p^2}{\pi^2}
\left(\frac23+\frac14\log\,2
\right)  +{\cal O}(p^4/\L_0^2)
\eeq
The gauge loop contribution \re{trvec}, can be evaluated in a similar way.
The presence of the square of the cutoff function makes the Feynman 
integrals more involved. 
By inserting the cutoff function \re{polcut} in \re{trvec} one 
obtains 
\beeq&&
{\cal G}_{\m\m}^{[1a]aa}+{\cal G}_{\m\m}^{[1b]aa}
=512 g^2\L_0^{8}\int_q\biggl\{\frac{6144\, p^2\L_0^4}
{((4\L_0^2+q^2)^4-(4\,\L_0^2)^4)((4\L_0^2+(q+p)^2)^4-(4\,\L_0^2)^4)}
\nonumber\\
&&\phantom{{\cal G}_{\m\m}^{[4a]aa}+{\cal G}_{\m\m}^{[4b]aa}}
-\frac{8}{q^2(4\L_0^2+q^2)(8\L_0^2+q^2)
(32\L_0^4+8\L_0^2q^2+q^4)}
\nonumber\\
&&\phantom{{\cal G}_{\m\m}^{[4a]aa}+{\cal G}_{\m\m}^{[4b]aa}}
+\frac{(2q+p)^2}{(8\L_0^2+q^2)}\biggl[
\frac{1}{(4\L_0^2+q^2)
(32\L_0^4+8\L_0^2q^2+q^4)(q+p)^2(4\L_0^2+(q+p)^2)}
\nonumber\\&&\phantom{{\cal G}_{\m\m}^{[4a]aa}+{\cal G}_{\m\m}^{[4b]aa}}
+\frac{8\L_0^2+q^2+(q+p)^2}{q^2(4\L_0^2+q^2)
(32\L_0^4+8\L_0^2q^2+q^4)(32\L_0^4+8\L_0^2(q+p)^2+(q+p)^4)}
\nonumber\\&&\phantom{{\cal G}_{\m\m}^{[4a]aa}+{\cal G}_{\m\m}^{[4b]aa}}
+\frac{1}{(32\L_0^4+8\L_0^2q^2+q^4)(q+p)^2(4\L_0^2+(q+p)^2)(8\L_0^2+(q+p)^2)}
\nonumber\\&&\phantom{{\cal G}_{\m\m}^{[4a]aa}+{\cal G}_{\m\m}^{[4b]aa}}
+\frac{1}{q^2(q+p)^2(4\L_0^2+q^2)(32\L_0^4+8q^2\L_0^2+q^4))}
\biggr]
\biggr\}\,.
\eeeq       
One can see that this integral is UV finite and that it vanishes for $p=0$. 
We only compute the contribution  originating from 
the first and the last terms  since the others do not 
generate $\log p^2/\L_0^2$, as can be seen by taking the $\UV$ limit 
in the integrand.
After applying the Feynman parameterization and integrating on the whole set 
of variables one obtains
\beq
-\frac{5p^2}{2\pi^2}\log\,\frac{p^2}{\L_0^2}+{\cal O}(p^2)\,.
\eeq
Adding the two contributions from the ghost and the gauge loop one obtains
\beq
{\cal G}(p/\L_0)=-\,\frac{g^2}{16\pi^2}\,\frac{22}{3} \log(p^2/\L_0^2)
+{\cal O}(1)
\,.
\eeq

\end{document}